\documentstyle[12pt,jsp,epsf,amssymb]{article}
\def\e{{\rm e}}
\def\o{{\rm o}}
\def\s{{\rm s}}
\def\l{{\rm l}}
\def\i{{\rm i}}
\def\d{{\rm d}}
\def\a{{\rm a}}
\def\t{{\rm t}}
\def\Tr{{\rm Tr}}
\def\dps{\displaystyle}

\begin{document}
\title{The Exact Solution of an Octagonal Rectangle Triangle Random Tiling}
\author{Jan de Gier \and Bernard Nienhuis}
\date{\normalsize\em Instituut voor Theoretische Fysica, Universiteit van
  Amsterdam, Valckenierstraat 65, 1018 XE Amsterdam, The Netherlands}
\maketitle
\thispagestyle{empty}
\begin{abstract}
We present a detailed calculation of the recently published exact solution of a
 random tiling model possessing an eight-fold symmetric phase. The
 solution is obtained using Bethe Ansatz and provides closed
 expressions for the entropy and phason elastic
 constants. Qualitatively, this model has the same features as the
 square-triangle random tiling model. We use the method of P. Kalugin,
 who solved the Bethe Ansatz equations for the square-triangle
 tiling, which were found by M. Widom.
\end{abstract}
\clearpage
\section{Introduction}
Random tiling models are ensembles of coverings of the plane, without
gaps or overlaps, with a set of rigid building blocks or tiles. The
most well-known application of such a model is perhaps the
anti-ferromagnetic Ising model on the triangular lattice,\cite{Blote:1982} of which the ground state configurations are described by
rhombus coverings of the plane. 

The discovery of quasi-crystals\cite{Shechtman:1984} revived the modeling
of structures by tiling models.\cite{Levine:1984} Quasi-crystals
are solid metallic alloys without periodicity but with long range
order (i.e. they have sharp diffraction peaks). Two competing
scenarios have been put forward to explain the origins of such
quasiperiodic ordering. One of them assumes a microscopic
Hamiltonian of which the ground state configuration is 
quasi-periodic. This would be something akin to the local matching
rules used by Penrose. These deterministic tilings are perfectly
quasiperiodic in the sense that their diffraction peaks have zero
width. As an alternative there is the random tiling picture, in which
one abandons the strict quasiperiodic long range order. These models
do have matching rules in the sense that tiles may neither overlap nor
have gaps between them, but these are not strong enough to enforce
a unique behavior at large distances. This then gives rise to an
ensemble of different tilings, or atom packings, inside a fixed
volume. Assuming that all these configurations are (nearly) degenerate
in energy such an ensemble has a nonzero entropy density. This entropy
can reduce the free energy and lower it w.r.t the crystalline phase,
which is supposed to be stable at zero temperature. This phenomenon
thus provides a simple explanation of the stability of the
quasi-crystalline phase.

Of special interest in two dimensional random tiling models are those
that fulfill the following constraints:
(i) each kind of tile has a unique atomic decoration, (ii) each member
of the random tiling ensemble is an allowed packing of atoms and (iii) each
arrangement of atoms forming an equally good packing belongs to the tiling
ensemble. These tiling models are called physical or
atomistic. Examples of these are the binary tenfold rhombus\cite{Widom:1989} and the twelve-fold square-triangle\cite{Collins:1964,Kawamura:1983} random 
tiling models. Recently Cockayne introduced a physical random tiling
model with an eight-fold symmetric phase which he doped the octagonal
analogue of the square-triangle tiling, despite of some distinctive
characteristics.\cite{Cockayne:1994} In this paper we show that the comparison
certainly holds concerning the exact solvability of the random tiling
model. We mean by this that, like for the square-triangle
tiling,\cite{Widom:1993,Kalugin:1994} it is possible to calculate the 
entropy exactly in part of the phase diagram. We obtain closed
expressions for the residual entropy as well as for the phason elastic
constants.\cite{Henley:1991} The main results of this paper have already been
published elsewhere,\cite{Gier:1996a} here we present the full calculation.

The paper is organized as follows: In section \ref{se:Tmat} we present
the tiling model and give a transfer matrix description to be able to
enumerate all possible tilings. In section \ref{se:Phas} we give a
brief review of phason strain concepts. The results of this section
will be used to identify our results and to obtain expressions for the
phason elastic constants. Section \ref{se:BA} describes the
diagonalization of the transfer matrix by Bethe Ansatz. The resulting
nonlinear coupled equations are solved in section
\ref{se:thermlim}. Finally we calculate the entropy density from this
solution in section \ref{se:entropy}.

\section{The square-hexagon tiling and the transfer matrix}
\label{se:Tmat}
\begin{figure}[h]
\centerline{\epsffile{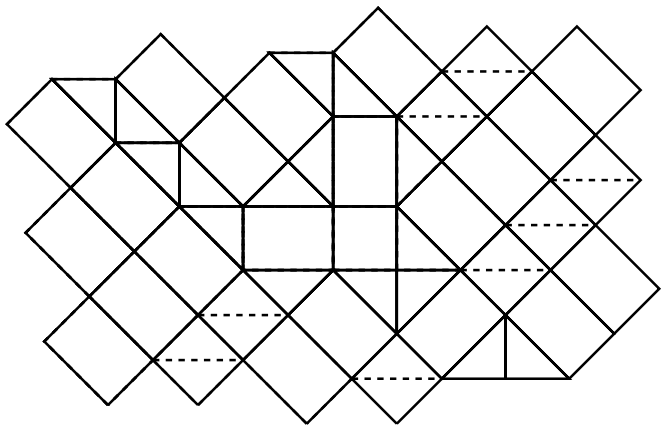}}
\caption{Patch of the tiling model. The diagonals of the squares are
  not drawn. Dashed lines indicate parts of the layer boundaries.}
\label{fig:Tflayerdecompo}
\end{figure}

The model that Cockayne described is a tiling of the plane by squares
and various hexagons. By placing discs of radius $1/2$ on each of the
vertices, one obtains a
physical packing. The hexagons can be thought of as being built out of
two isosceles triangles and several rectangles depending on the length
of the hexagon. The rectangles have sides 1 and $\sqrt{2}$. Likewise, by
drawing one of the two diagonals, the squares can be viewed as two
triangles. The square-hexagon tiling thus is
equivalent to the ensemble of vertex configurations of the
rectangle-triangle tiling. The latter ensemble in turn is equivalent
to the rectangle-triangle random tiling ensemble (see
Fig.~\ref{fig:Tflayerdecompo}) provided that the two ways of forming a
square out of two triangles are counted as one.\cite{Gier:1996a}

\begin{figure}[h]
\centerline{\epsffile{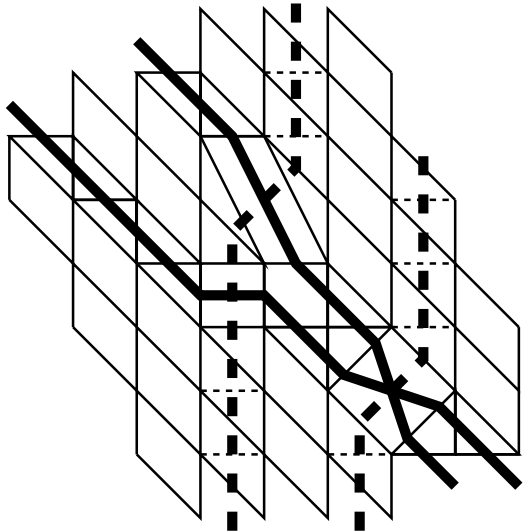}}
\caption{Corresponding patch on the
  lattice. Bold solid lines are domain walls of horizontal short
  edges, referred to as s-walls in the text. Bold dashed lines
  represent the l-walls of horizontal long edges.}
\label{fig:Tflattice}
\end{figure}

Like for the square-triangle tiling we make use of the transfer
matrix to calculate thermodynamic quantities like the entropy. The
transfer matrix {\bf T} is introduced by decomposing the
tiling into layers. Different layers are bounded by the short
horizontal edges, the horizontal diagonals of the squares and the
almost horizontal diagonals of the $\pm \pi/4$ tilted rectangles. Here,
$\pm$ refers to the tilt of the 
short edge of the rectangle w.r.t the horizontal axis. In addition,
the layer boundaries cut the
triangles and rectangles with a vertical long edge in half. The
horizontal diagonals of the squares are denoted by the dashed lines in
Fig.~\ref{fig:Tflayerdecompo}. In a similar fashion the tiling can
be decomposed in `columns'. Different columns are separated by the
short + edge, the vertical short edge and the almost vertical diagonal
of the $-$ tilted rectangle. The rectangles with a vertical long edge
are cut by a column edge from the lower left to the upper right corner.
In this way the tiles are deformed in such a
way that the vertices of the tiling form a subset of those of the
square lattice, see Fig.~\ref{fig:Tflattice}. 
\\
On this regular lattice the
definition of the transfer matrix {\bf T} is obvious:
a matrix element $T_{ij}$ is 0 if layer $j$ can not be followed by
layer $i$. Otherwise it is given by the statistical weight of layer
$i$. In this paper we shall adopt the convention that ${\bf T}$ acts
in the downward direction.

In the following we will call
the tilted rectangles $R_{\pm}$ and the rectangles with the short and long
horizontal edge by $R_\s$ and $R_\l$ respectively. 
Because different tiles of the original tiling are mapped onto the same
shapes on the square lattice, we have to decorate the new configurations. This
is done with bold dashed and solid lines, see
Fig.~\ref{fig:Tflattice}. Thus it is apparent that the horizontal short
and long edges of the original tiling form domain walls between
regions of $R_+$, which we denote by
type s and l respectively. If we think of the vertical coordinate
as a time coordinate (increasing in the downward direction), the transfer matrix can be seen as an evolution
operator for these domain walls. The transfer matrix {\bf T} is block diagonal
with sectors parameterized by the numbers of domain walls, which are
conserved under the action of {\bf T}. 
\\
Between two layer edges on the square
lattice, the s-walls step one unit to the right and the l-walls do
not move. When two walls cross, the s-wall may either jump over the
l-wall moving two places to the right and thereby creating a
rectangle $R_\l$, or, over {\em two} layers, the walls may
exchange place creating a rectangle $\rm R_\s$. In the latter case the
crossing therefore is completed after application of the
transfer matrix twice. It may also happen that two walls of type s
and one of type l cross simultaneously over two layers. The l-wall
and the s-wall nearest to it then exchange place, while the 
second s-wall jumps over both these walls moving three places to the
right creating a rectangle $R_-$.   

We can express the tile densities in terms of the domain wall
densities. We shall denote the horizontal size of the tiling by $L$ and the
corresponding system size of the lattice model by $N$. Let
$\Delta_{\l\s}=n_{R_\l}-n_{R_\s}=0$, i.e. both types of collisions of
two domain walls occur with the same frequency, and let $2n_\s$ and
$n_\l$ be the number of s- and l-walls. Apply the 
transfer matrix $p=2N-2n_\s$ times on some initial configuration of
domain walls at $t=0$ on the lattice, and suppose that both types of collisions
occur for every {\em pair} of s- and l-walls. The final state at
$t=2N-2n_\s$ will then be the same configuration of domain walls as the
initial one shifted by $-2n_\s$. Counting the total number of tiles in
this patch generated by {\bf T}, we find the following relations
between the wall densities and the rectangle densities per layer:
\begin{eqnarray}
p(n_{R_\l}+n_{R_\s}+2n_{R_-})=2p(n_{R_\s}+n_{R_-})=4n_\s
n_\l.\label{eq:TfRsR-}\\ 
\Delta_{\pm}\equiv n_{R_+}-n_{R_-}=N-2n_\s-n_\l.\label{eq:Tfepsdef}
\end{eqnarray}
Similar expressions can be obtained for the different triangle densities.
The total number of rectangles and of triangles per layer can then be
calculated to be: 
\begin{equation}
\begin{array}{l}
\dps n_{\rm rect} = N-2n_\s-n_\l+4n_\s n_\l/p.\\
n_{\rm tri} = 2(2n_\s+n_\l)-12n_\s n_\l/p.
\end{array}
\label{eq:Tftiledens}\end{equation}
The tile densities that belong to the quasi-crystalline phase are
$n_{\rm rect}/N=6-4\sqrt{2},\;n_{\rm tri}/N=12\sqrt{2}-16$, corresponding to an
area fraction of triangles $\alpha_t=1/2$. 
As a function of the domain wall densities, the model displays
two incommensurate phases. A four-fold symmetric phase is formed in the high
density region, $\alpha_t>1/2$, where the triangles form octagonal and square
cells bounded by domain walls consisting of rectangles. There is a two-fold
symmetric phase in the low density region where the rectangles from
rectangular cells bounded by the domain walls consisting of triangles.

As already mentioned above, the quantities $n_\s$ and $n_\l$ are
conserved by the action of the transfer matrix {\bf T}. To control the
average value of $\Delta_{\l\s}$, the tiles $R_\s$
and $R_\l$ are given a weight $\exp(-\phi)$ and $\exp(\phi)$
respectively. Furthermore, as the tiles $R_\s$ and $R_-$ in the lattice
representation have an area that is twice that of the other two
transformed rectangles, we have to introduce a chemical potential for
them to compensate for this asymmetry. The tiles $R_\s$ and $R_-$
therefore get an extra weight $\exp(\eta)$.  

The free energy per layer of the lattice model is given by the logarithm of the
largest eigenvalue of ${\bf T}$:
\begin{eqnarray}
f\left(n_\s,n_\l,\phi\right)&\;=\;&-\log\Lambda\nonumber\\
&\;=\;&-N\sigma_N-\phi\Delta_{\l\s}-\eta\left(n_{R_\s}+n_{R_-}\right),\label{eq:Tffren}
\end{eqnarray}
where $\sigma_N$ is the entropy per site of the lattice model. In
section~\ref{se:BA} will be shown that we can actually diagonalize
${\bf T}$ using coordinate Bethe Ansatz. 

\section{Phason strain}
\label{se:Phas}
In this section we will briefly review some of the phason strain
concepts. Details can be found in a review on random
tiling concepts by Henley.\cite{Henley:1991}.

The position of every vertex in the tiling is of the form:
\begin{equation}\begin{array}{l}
\dps {\mathbf r} = \sum_{j=0}^3 n_j {\mathbf e}^{||}_j,\;\;n_j\in \mathbb{N}.\\
\dps {\mathbf e}^{||}_j = (\cos \pi j/4, \sin \pi j/4).
\end{array}\end{equation}
The tiling can therefore naturally be embedded in a four-dimensional
hyper-cubic lattice using
\begin{equation}\begin{array}{l}
\dps ({\bf r},{\bf h}) = \sum_{j=0}^3 n_j
({\bf e}^{||}_j,{\bf e}^{\perp}_j),\\
\dps {\bf e}^{\perp}_j = (\cos 5\pi j/4, \sin 5\pi j/4).
\end{array}\end{equation}
The vectors ${\bf e}^{\perp}_j$ are chosen such that
${\bf e}_j=({\bf e}^{||}_j,{\bf e}^{\perp}_j)$ are
orthogonal. The continuous height field, or representative surface,
${\bf h}({\bf r})$ is as horizontal as possible for a perfect
quasi-crystalline tiling. A generic member of the random tiling
ensemble thus is described by fluctuations around this flat
surface. By integrating out the short-wavelength fluctuations one gets
a smoothed function ${\bf{\tilde{h}}}({\bf r})$, the phason field, that
varies on length scales much larger than the tile edge length. The
phason strain tensor then is defined by linearization:
\begin{equation}
{\bf E} = \nabla_{{\bf r}}{\bf \tilde{h}}({\bf r}).
\end{equation}
The macroscopic quasi-crystalline state corresponds to ${\bf E}=0$. The
random tiling hypotheses now states that the entropy density
$\sigma_a({\bf E})=S({\bf E})/A$ has its maximum at ${\bf E}=0$ and
that it is quadratic in ${\bf E}$ near ${\bf E}=0$ for any maximally
random tiling. Like the square-triangle tiling, the octagonal tiling
has the irrotational property.\cite{Oxborrow:1993} This implies that the
most general quadratic form of the entropy density $\sigma_a$
consistent with 8-fold rotational symmetry is
\begin{equation}
\sigma_\a = \sigma_{\a,0}-{1 \over 2}K_{\mu}\left({\rm Tr}{\bf E}\right)^2+{1
  \over 2}K_{\xi}\det{\bf E}+{\cal{O}}\left({\bf E}^3\right)\label{eq:Psrthyp}
\end{equation}

We denote the deviations of the quasi-crystalline densities by
\begin{equation}\begin{array}{l}
\delta_{\l\s} = n_\l\sqrt{2}-2n_s.\\
\Delta_{\pm} = N-n_\l-2n_\s.
\end{array}\label{eq:Psdeltas}\end{equation}
The quadratic forms in (\ref{eq:Psrthyp}) can be expressed in these, using
$L=n_\l \sqrt{2}+2n_\s$
\begin{equation}\begin{array}{rcl}
\dps \left(\Tr{\bf E}\right)^2 &=& \dps {1 \over L^2}\left(2\delta_{\l\s}-
  \Delta_{\pm}(2-\sqrt{2}) -
  \Delta_{\l\s}(1+\sqrt{2})\right)^2.\\[2mm]  
\dps \det{\bf
  E} &=& \dps {1\over L^2}\left(\vphantom{1\over 2}\delta_{\l\s}^2-(2-
\sqrt{2})\Delta_{\pm}\delta_{\l\s} - 2\Delta_{\pm}^2\sqrt{2}\right.\\
&&\dps \hphantom{{1\over L^2}\left(\vphantom{1\over 2}\right.}\left. -
(1+\sqrt{2})\Delta_{\l\s}(\delta_{\l\s}-(1-{1\over
  2}\sqrt{2})\Delta_{\pm})\right). 
\end{array}\label{eq:Psinvariants} \end{equation}
The conditions on the elastic constants for ${\bf E}=0$ to be a local
maximum of $\sigma_a$ are:
\begin{equation}
K_{\mu}>0,\;\;K_{\xi}>0,\;\;4K_{\mu}-K_{\xi}>0.
\end{equation}
  
\section{Bethe Ansatz}
\label{se:BA}
It is possible to diagonalize the transfer matrix by making an Ansatz
for its eigenvectors. Details of this so-called Bethe Ansatz are given
in Appendix~\ref{ap:BA}. A technical point we mention here is the fact that the
s-walls split up in two different species, odd and even ones, both
of which are conserved under the action of ${\bf T}$. Their
densities are given by $n_\o$ and $n_\e$ respectively, with
$n_\o+n_\e=2n_\s$. The
eigenvectors and eigenvalue can be expressed in three sets of complex
numbers corresponding to the three different kinds of domain wall. The
eigenvalue is given by
\begin{equation}
\Lambda = \prod_{i=1}^{n_\o} u_i \prod_{j=1}^{n_\e}
v_j,\label{eq:BAeigval}
\end{equation}
if the numbers $u_i$ and $v_j$ satisfy the following equations 
\begin{equation}\begin{array}{rcl}
\dps u_i^{N} &=& \dps (-)^{n_\o-1}\prod_{k=1}^{n_\l}(\e^\phi
  u_i+\e^{\eta-\phi} u_i^{-1} w_k^{-1}).\\[6mm]
\dps v_j^{N} &=& \dps (-)^{n_\e-1}\prod_{k=1}^{n_\l}(\e^\phi
  v_j+\e^{\eta-\phi} v_j^{-1} w_k^{-1}).\\[6mm]
\dps w_k^{-N} &=& \dps (-)^{n_\l-1} \prod_{i=1}^{n_\o}(\e^\phi
  u_i+\e^{\eta-\phi} u_i^{-1} w_k^{-1}) \prod_{j=1}^{n_\e}(\e^\phi
  v_j+\e^{\eta-\phi} v_j^{-1} w_k^{-1}).
\end{array}\label{eq:BABAE}\end{equation}
The equations (\ref{eq:BABAE}) are the Bethe Ansatz equations
(BAE). Using the substitution 
\begin{equation}
\xi_{\e,j}=v_j^2,\;\;\xi_{\o,i}=u_i^2,\;\;\psi_j=-w_j^{-1}\e^{\eta-2\phi},
\end{equation}
their logarithmic version can be written as
\begin{equation}\begin{array}{l}
\dps \frac{N+n_l}{2}F_\s(\xi_{\o,i}) = 0\; (\bmod{2\pi \i}),\\[2mm]
\dps \frac{N+n_l}{2}F_\s(\xi_{\e,i}) = 0\; (\bmod{2\pi \i}),\\[2mm]
\dps N F_\l(\psi_j) = 0\;(\bmod{2\pi \i}),
\end{array}\label{eq:BAlogba}\end{equation}
where we have defined the following functions
\begin{eqnarray}
F_\s(z) &=& \log (z) - \frac{2}{N+n_\l} \sum_{j=1}^{n_\l}
\log\left(z-\psi_j\right) - \frac{2n_\l}{N+n_\l}\phi. \label{eq:BAFsdef}\\
F_\l(z) &=& \log (-z) - \frac{1}{N} \sum_{i=1}^{n_\o}
\log\left(\xi_{\o,i}-z\right) - \frac{1}{N} \sum_{i=1}^{n_\e}
\log\left(\xi_{\e,i}-z\right) \nonumber\\
&&{}+\frac{1}{N}\log\left(\Lambda\right)+\frac{2N-n_\o-n_\e}{N}\phi-\eta.
 \label{eq:BAFldef}
\end{eqnarray}
The expression for the eigenvalue $\Lambda$ of the transfer matrix
(\ref{eq:BAeigval}) becomes
\begin{equation}
2\log\left(\Lambda\right)=\sum_{i=1}^{n_\o}\log\left(\xi_{\o,i}\right)+\sum_{i=1}^{n_\e}\log\left(\xi_{\e,i}\right).\label{eq:BAlambda}
\end{equation}
The Ansatz (\ref{eq:apBAvector}) together with the BAE (\ref{eq:BABAE})
correctly give the eigenvector with eigenvalue (\ref{eq:BAlambda}) of the
transfer matrix for {\em almost} all densities. The description fails
for the two states in which 
the lattice is filled completely with either s- or l-walls. In
these cases it is possible to have 'virtual' l- or s-walls
respectively, running around the cylinder. 

\section{Thermodynamic limit}
\label{se:thermlim}
\begin{figure}[h]
\centerline{\epsffile{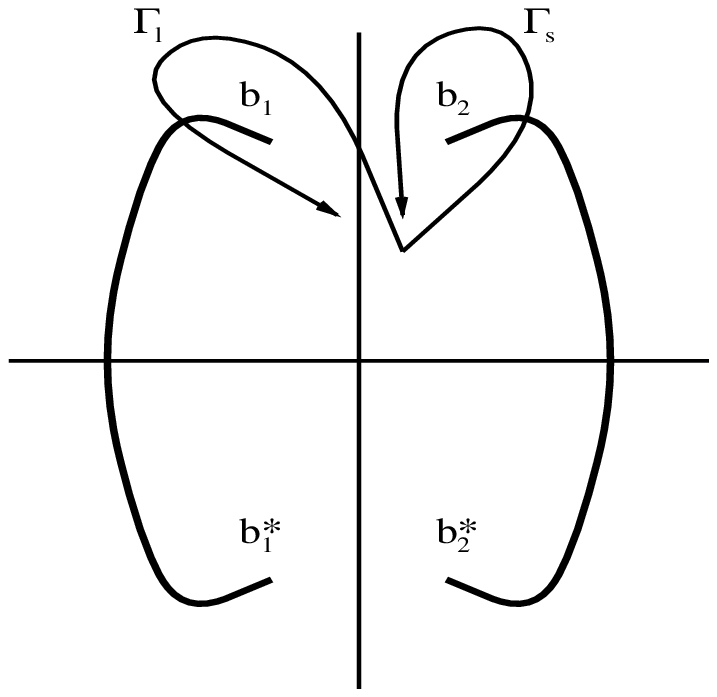}}
\caption{BAE curves.}
\label{fig:Tlcurves}
\end{figure}

It is clear that for the largest eigenvalue we must have
$n_\o=n_\e$. In that case the solution for $\xi_{\o,i}$ and
$\xi_{\e,i}$ must be the same. If not, the largest eigenvalue would be
degenerate which cannot be the case by the Perron-Frobenius theorem.
We therefore from now on omit the subscripts o and e and we set
$n_\o=n_\e=n_\s$, as well as $\xi_{\o,i}=\xi_{\e,i}=\xi_i$. 
The roots $\xi_i$ and $\psi_j$ approach two curves in the complex
plane (see Fig.~\ref{fig:Tlcurves}). We shall refer to these curves as
$\Xi$ and $\Psi$ respectively.\\

From numerical calculations we note that there are no holes in
the solution for the largest eigenvalue. 
Under the assumption that there are no holes in the thermodynamic
limit of the solution we infer from the BAE (\ref{eq:BAlogba}) that for $(N\rightarrow\infty)$
\begin{eqnarray}
{N+n_\l\over2}f_\s(\xi_k)(\xi_{k+1}-\xi_k)&=&2\pi\i,\label{eq:TlXidef}\\
Nf_\l(\psi_k)(\psi_{k+1}-\psi_k)&=&2\pi\i\label{eq:TlPsidef},
\end{eqnarray}
where $f_\s$ and $f_\l$ are the derivatives of the functions $F_\s$ and
$F_\l$ respectively.
These equations allow us to convert the sums in (\ref{eq:BAFsdef}) and
(\ref{eq:BAFldef}) into contour-integrals. By introducing the densities
$Q_\l=n_\l/N$ and $Q_\s=n_\s/N$ we obtain the following relations between
$f_\s$ and $f_\l$ in the thermodynamic limit:
\begin{eqnarray}
f_\s(z)&=&{1\over z}-{2\over 1+Q_\l}{1\over
2\pi\i}\int_{b_1^*}^{b_1}{f_\l(\psi)\over z-\psi}d\psi,\label{eq:Tlfsint}\\
f_\l(z)&=&{1\over z}+{1+Q_\l\over
2\pi\i}\int_{b_2^*}^{b_2}{f_\s(\xi)\over z-\xi}d\xi.\label{eq:Tlflint}
\end{eqnarray}
The points $b_1,b_1^*,b_2$ and $b_2^*$ are the thermodynamic
limit points of $\psi_{1},\psi_{n_\l},\xi_{n_\s}$ and $\xi_{1}$.
From the form of equation (\ref{eq:Tlfsint}) and (\ref{eq:Tlflint}) we see that
$f_\s(z)$ is analytic along the contour $\Gamma_\s$, though $f_\l(z)$ is
not. The function $f_\l(z)$ differs on both sides of the curve $\Xi$ by
the amount
\begin{equation}
{1+Q_\l\over 2\pi\i}\oint_{|\xi-z_*|=\epsilon}{f_\s(\xi)d\xi\over
  z_*-\xi}=-(1+Q_\l)f_\s(z_*),
\end{equation}
where $z_*$ is the point where $\Gamma_\s$ crosses the curve $\Xi$.
The analytic continuation of the function $f_\l(z)$ along $\Gamma_s$ therefore
equals the function $f_\l(z)+(1+Q_\l)f_\s(z)$. A similar argument can be
used for the analytic properties of the functions $f_\s$ and $f_\l$
along the contour $\Gamma_\l$. The analytic properties of a linear combination
$G(z)=a_\s f_\s(z)+a_\l f_\l(z)$ along the two contours are described by the so
called monodromy operators $\Gamma_\s$ and $\Gamma_\l$:
\begin{eqnarray}
\Gamma_\s: \left(\begin{array}{c} a_\s\\a_\l \end{array}\right)\mapsto\left(\begin{array}{cc} 1 &
1+Q_\l\\ 0 & 1\end{array}\right)\left(\begin{array}{c} a_\s\\a_\l \end{array}\right).\\
\Gamma_\l: \left(\begin{array}{c} a_\s\\a_\l \end{array}\right)\mapsto\left(\begin{array}{cc} 1 &
0\\ -{2\over 1+Q_\l} & 1\end{array}\right)\left(\begin{array}{c} a_\s\\a_\l \end{array}\right).
\end{eqnarray}
If the two curves close, i.e. they have the same limit point
$b_1=b_2=b$, it is not possible to close the contours $\Gamma_\s$ and
$\Gamma_\l$. The only nontrivial closed curve is one that crosses both
the curves $\Xi$ and $\Psi$, so that we are left with only one
monodromy operator:  
\begin{equation}
\Gamma_\l\Gamma_\s=\left(\begin{array}{cc} 1 & 1+Q_\l \\ -{2\over 1+Q_\l} & -1
\end{array}\right),\label{eq:Tlmonop}
\end{equation}
which has the property that $\left(\Gamma_\l\Gamma_\s\right)^4=1$.
This means that in this case the function $G(z)$ is a single-valued
function of the parameter
\begin{equation} t(z)=\left({zb^{-1}-1\over 1-b^{*^{-1}}z}\right)^{1\over
4},\;\;z(t)=b{1+t^4\over 1+bb^{*^{-1}}t^4}.\label{eq:Tltdef}
\end{equation}
From now on we will write $b=\i|b|\e^{\i\gamma}$. 
If we look at the orbit of $f_\l(z)$ under $\Gamma_\l\Gamma_\s$ we can
see that $f_\l(z)$, $(1+Q_\l)f_\s(z)-f_\l(z)$, $-f_\l(z)$ and
$-(1+Q_\l)f_\s(z)+f_\l(z)$ belong to different sheets of the Riemann
surface of the same function $G$, where $G$ is given by (Appendix
\ref{ap:calcG}):
\begin{eqnarray}
G&=&\left(\left(\e^{-\i\pi/4}t+\e^{\i\pi/4}t^{-1}\right)-Q_l\left(\e^{\i\pi/4}t+\e^{-\i\pi/4}t^{-1}\right)\right){1\over 2z(t)}.\label{eq:TlGt}
\end{eqnarray}

\section{Entropy}
\label{se:entropy}
The knowledge of the function $G$ allows us to calculate the
eigenvalue from the BAE (\ref{eq:BAlogba}), see
appendix \ref{ap:calclamb}. Equations (\ref{eq:TfRsR-}) and (\ref{eq:Tffren})
show that the entropy per site for $\Delta_{\l\s}=0$
is given by: 
\begin{equation}
\sigma_N = N^{-1}\log\Lambda-{Q_\s Q_\l\over 1-Q_\s}\eta.
\end{equation}
This expression is precisely given by the solution (\ref{eq:apcleigval})
of the BAE. 

At the QC-point, $\gamma=0,C=0,\phi=0$ (for the definition of $C$, see
appendix \ref{ap:calcG}) and the domain wall densities
given by (\ref{eq:apCGQs}) and (\ref{eq:apCGQl}) become
\begin{equation}
Q_\l = \sqrt{2}-1,\qquad 2Q_\s = 2-\sqrt{2}.
\end{equation}
This implies via (\ref{eq:Tftiledens}) that the area per site $A/N$ on the QC-point is given by
\begin{equation}
A/N=12\sqrt{2}-16,\qquad Q_{\rm rect}=6-4\sqrt{2},\qquad Q_{\rm
  tri}=12\sqrt{2}-16. 
\end{equation}
Therefore the entropy per area at the QC-point is given by
\begin{eqnarray}
\sigma_{\a,0}=(A/N)^{-1}\sigma_N &=& {1+\sqrt{2}\over 2\sqrt{2}}\left(\log
4-\sqrt{2}\log(1+\sqrt{2})\right)\\[2mm]
&\approx& 0.1193642186\ldots\nonumber
\end{eqnarray}

Away from the QC-point, there are two incommensurate phases. There is a two-fold
degenerate four-fold symmetric phase in which one type of triangle (say
the up and down pointing ones) dominates while the other type is
suppressed. The triangles form rectangular and octagonal cells bounded
by domain walls consisting of rectangles. Each orientation of the
rectangles occurs equally often.  
A four-fold degenerate two-fold symmetric phase is present, if one type of
rectangle dominates (say $R_+$, as in Fig.~\ref{fig:Tflayerdecompo})
forming rectangular cells bounded by domain walls consisting of
triangles. The 90 degrees rotated rectangle (here $R_-$) is suppressed
in this phase and the other two types of rectangle occur equally
often, as well as both types of triangle.

Two independent parameters describe the deviations from the QC-point into
these phases. The four-fold phase is described by the parameter
$\gamma$ which measures the difference in the densities of the two
types of triangle. This phase also requires
$\Delta_{ls}=\partial\log\Lambda/\partial\phi=0$. The two-fold phase
is described by $\epsilon=\Delta_{\pm}/N=1-Q_\l-2Q_\s$. This parameter
measures the difference in densities of the $\pm$ tilted rectangles,
see equation (\ref{eq:Tfepsdef}). In this phase $\phi=0$ and
$\partial\log\Lambda/\partial\phi=0$.  

\subsection{Two-fold symmetric phase}
\label{sse:E2-foldphase}
In the two-fold phase $\gamma=\Delta_{\l\s}=\phi=0$ and the densities
(\ref{eq:apCGQs}) and (\ref{eq:apCGQl}) become
\begin{eqnarray}
Q_\l&=&(\sqrt{2}-1)(1+4\Re(C)+4\Im(C)).\nonumber\\
2Q_\s&=&2-\sqrt{2}-2\sqrt{2}\Re(C)+2(2-\sqrt{2})\Im(C).
\end{eqnarray}
Using the fact that $\delta_{\l\s}/N=Q_\l\sqrt{2}-2Q_\s=0$ in the
two-fold phase, $C$ can be expressed in $\epsilon$:
\begin{equation}
\Re(C)={\sqrt{2}-1\over
  4\sqrt{2}}\epsilon,\;\;\Im(C)=-{2\sqrt{2}-1\over 4\sqrt{2}}\epsilon. 
\end{equation}
Since a nonzero value of $\epsilon$ implies that $C\neq 0$ it thus not only
affects the value of the residues, or equivalently the densities, but
also the topology of the Riemann
surface of $G$ (see appendix \ref{ap:calcG}). The formula
(\ref{eq:apCGG}) gives the lowest order corrections compatible with
the modifications to the residues using the above value for $C$. It does
not however take into account the changes in the topology. Those
corrections become comparable with (\ref{eq:apCGG}) when
$|b_1-b_2|^{1/2}\sim \epsilon$ (see the end of appendix
\ref{ap:calcG}). Since we want to know the corrections 
to the entropy density we need to know how small changes in the topology
modify the integrals in appendix \ref{ap:calclamb}. These corrections
arise from the contributions of the poles at $z=0$ and $z=\infty$
on other parts of the Riemann surface, which were previously
disconnected. Since the net total residue on each part of the surface
is zero, the lowest order correction to the integrals in appendix
\ref{ap:calclamb} will be of the order of $|b_1-b_2|^2\sim
\epsilon^4$. The entropy density 
is linear in the integrals of appendix \ref{ap:calclamb}, the lowest order
corrections to the entropy can therefore be calculated using the modifications
to the residues only.
The entropy per area up to second order in $\epsilon$ then becomes:
\begin{equation}
\sigma_\a=\sigma_{\a,0}-\epsilon^2{1+\sqrt{2}\over 8}\left(\log{4}+\sqrt{2}\log(1+\sqrt{2})\right).
\end{equation}
Comparing this expansion with the expression for $\sigma_\a$ in terms
of the phason elasticity (\ref{eq:Psrthyp}) one obtains for the elastic constants:
\begin{equation}
K_{\mu}\left(\sqrt{2}-1\right)^2+K_{\xi}\sqrt{2}=\left(\sqrt{2}-1\right)\left(\log
4+\sqrt{2}\log\left(1+\sqrt{2}\right)\right).\label{eq:Eeps_elconst}
\end{equation}

\subsection{Four-fold symmetric phase}
In the four-fold symmetric phase $\epsilon=0$. In
this case the entropy throughout the entire four-fold phase can be
calculated exactly. First we note that
$\Delta_{\l\s}=\partial\Lambda/\partial\phi$, which can be
calculated analogously to the lowest order corrections in
$\epsilon$. It is then found that 
\begin{equation}
\frac{\partial\Lambda}{\partial\phi}\sim \left.\frac{\partial \Lambda}{\partial C}\right|_{C=0} = 0.
\end{equation}
The densities in this case are 
\begin{eqnarray}
Q_\l&=&\left(1+\sqrt{2}\cos{\gamma\over
  2}\right)^{-1}\left(1-\sqrt{2}\sin{\gamma\over 2}\right).\\
2Q_\s&=&\left(1+\sqrt{2}\cos{\gamma\over
  2}\right)^{-1}\left(\sqrt{2}\cos{\gamma\over 2}+\sqrt{2}\sin{\gamma\over
  2}\right). 
\end{eqnarray}
and the layer separation is
\begin{equation}
A/N=1-(\sqrt{2}-1)^2{\sqrt{2}\cos{\gamma/ 2}-1\over
  \sqrt{2}\cos{\gamma/ 2}+1}.
\end{equation}
The area fraction of triangles in terms of $\gamma$ then can be
written as
\begin{equation}
\alpha_\t = {Q_{\rm tri}{1\over 2}\over
  A/N}=(\sqrt{2}+1){\sqrt{2}-\cos{\gamma/ 2}\over
  1+\cos{\gamma/ 2}},
\end{equation}
so that the four-fold symmetric phase corresponds to
$1/2\leq\alpha_\t\leq 1$. The entropy per area in this regime is found
to have the following form
\begin{eqnarray}
\sigma_\a &=& {2+\sqrt{2}\over 4\cos^2\gamma/4} \left(\log \left({4\over\cos\gamma}\right)\right.\nonumber\\
&&{}+\cos\left({\pi\over 4}+{\gamma\over 2}\right)\log\tan
\left({\pi\over 8}+{\gamma\over 4}\right)\nonumber\\
&&\left.{}+\cos\left({\pi\over 4}-{\gamma\over 2}\right)\log\tan
\left({\pi\over 8}-{\gamma\over 4}\right)\right).
\end{eqnarray}
\begin{figure}[h]
\centerline{\epsffile{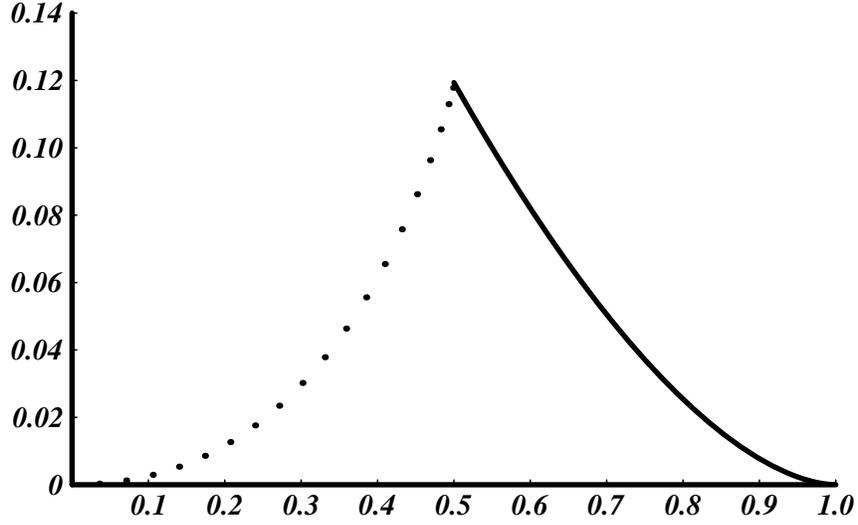}}
\caption{Entropy as a function of $\alpha_\t$. Solid line corresponds
  to the exact solution. Dots are numerical results for N=198.}
\label{fig:Eentropy}
\end{figure}

Expanding $\sigma_\a$ up to lowest order in $\gamma$ yields
\begin{equation}
\sigma_\a=\sigma_{\a,0}-\gamma^2{1+\sqrt{2}\over 32\sqrt{2}}\left(4-\log
4-\sqrt{2}\log\left(1+\sqrt{2}\right)\right).
\end{equation}
To compare this with the phason strain description we note that in
lowest order in $\gamma$
\begin{equation}
\delta_{\l\s}/L = -{\gamma \left(\sqrt{2}+1\right)\over 4}.
\end{equation}
Using (\ref{eq:Psinvariants}) and (\ref{eq:Psrthyp}) we then find that
\begin{eqnarray}
4K_{\mu}-K_{\xi}&=&{\sqrt{2}-1\over\sqrt{2}}\left(4-\log
4-\sqrt{2}\log (1+\sqrt{2})\right)\nonumber\\
&\approx & 0.4004597643\dots\label{eq:Egam_elconst}
\end{eqnarray}
Combining (\ref{eq:Eeps_elconst}) and (\ref{eq:Egam_elconst}) one obtains the phason elastic constants:
\begin{eqnarray}
K_{\mu}&=&4 (\sqrt{2}-1)^3\approx0.2842712475\dots\\
K_{\xi}&=&{\sqrt{2}-1\over\sqrt{2}}\left(\log 4+\sqrt{2}\log
(1+\sqrt{2})-4 (\sqrt{2}-1)^4\right)\\
&\approx & 0.7366252255\dots\nonumber
\end{eqnarray}
Qualitatively, this octagonal tiling displays the same behavior as
the square-triangle tiling.\cite{Widom:1993} At $\alpha_\t={1\over 2}$ the
model undergoes a so-called anomalous first order transition. This
means that at fixed density the derivative of the entropy w.r.t. to
the density, i.e. the pressure, has a jump, see
Fig.~\ref{fig:Eentropy}. This then provides a mechanism for locking on the
quasi-crystal phase over a range of pressures. 
\section{Conclusion}
In this paper we successfully applied Bethe Ansatz to the square
hexagon random tiling model. This model is a
physical,\cite{Oxborrow:1993} or atomistic, model of a  
quasi-crystal with eight-fold rotational symmetry. Amongst other things
its solution answers the question posed in\cite{Cockayne:1994} if the
random tiling hypothesis would hold for this model. It remains unclear
however, what the precise conditions for solvability of these tiling
models are. In this respect, the irrotationality constraint, which is
satisfied by both the square-triangle and the square-hexagon tiling,
might be an important guide. Such a constraint, not
necessarily irrotationality, which reduces the degrees of freedom
probably has some influence on the integrability. The tenfold tiling
by rectangles and triangles\cite{Oxborrow:1994} however does admit a Bethe
Ansatz. We know not of any constraint for this tiling, nor have we
solved the Bethe Ansatz equations. In the future we hope to present
some numerical results for this tiling.
\section*{Acknowledgment}
This research was supported by FOM, which is part of the Dutch Foundation for
Scientific Research NWO. 

\appendix
\section{Bethe Ansatz}
\label{ap:BA}
In this section we show how to set up an Ansatz for the eigenvectors
of the transfer matrix {\bf T} to solve the eigenvalue problem. The
conserved quantities make {\bf T} block-diagonal. We can make use of
this fact by diagonalizing {\bf T} in each sector
separately. It is helpful to start with the easiest sectors
$2n_\s=0,1$, $n_\l=0,1$. Later we generalize the method to arbitrary
values of $2n_\s$, $n_\l$.

\subsection{Domain walls of one kind only}
The sectors where either $n_\l=0$ or $n_\s=0$ are very simple. E.g., in a
sector with $n_\l=0$, ${\bf T}$ just shifts all s-walls one step to
the right. If we assign a wave-vector $p_i$ to every s-wall and
denote the coordinate of the $i$th s-wall by $x_i$, the eigenvectors
$\psi(\{x_i\})$ of {\bf T} in these sectors can be written as a product of
plane waves:  
\begin{equation}
\psi(\{x_i\}) = \prod_{i=1}^{2n_\s}\e^{-\i p_ix_i}.
\end{equation}
The eigenvalue in terms of the wave-vectors is:
\begin{equation}
\Lambda = \prod_{i=1}^{2n_\s} \e^{\i p_i}.
\end{equation}
When using periodic boundary conditions the wave-vectors have to obey the
following equation:
\begin{equation}
\e^{\i Np_i} = 1.
\end{equation}
These equations are easily solved to give the spectrum of {\bf T} in
these sectors. For later convenience we will fix some notation. The
wave-vectors for the l-walls will be called $q_j$ and in the
following we will make frequently use of the exponentiated momenta
$u_i=\exp(\i p_i)$ and $w_j=\exp(\i q_j)$.
\subsection{The sector $n_\l=1$, $2n_\s=1$.}
\begin{figure}[h]
\centerline{\epsffile{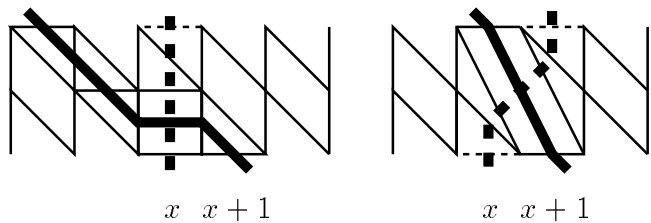}}
\caption{Crossings of two domain walls.}
\label{fig:2partcross}
\end{figure}

The simplest non-trivial sector is that with one s- and one
l-wall, see Fig.~\ref{fig:2partcross}. Let $z$ be the coordinate of the
l-wall, the eigenvalue equations then read:
\begin{eqnarray}
\Lambda \psi(x;z)&=&\psi(x-1;z),\;x > z+1.\label{eq:apBAeigeqonea}\\
\Lambda \psi(x;z)&=&\psi(x-1;z),\;x < z.\label{eq:apBAeigeqoneb}\\
\Lambda \psi(x+1;x)&=&\e^{(\eta-\phi)/2}\psi(x;x)+\e^{\phi}\psi(x-1;x).\label{eq:apBAeigeqonec}\\
\Lambda \psi(x;x)&=&\e^{(\eta-\phi)/2}\psi(x;x+1).\label{eq:apBAeigeqoned}
\end{eqnarray}
We do not have to specify exactly the positions of the
walls in the intermediate state $\psi(x;x)$ when the walls cross via
$R_\s$. We only denote this state by $\psi(x;x)$ which we can express
in $\psi(x;x+1)$ with equation (\ref{eq:apBAeigeqoned}). 
\\
We can solve this eigenvalue problem by making the following plane
wave Ansatz for the eigenfunction 
\begin{eqnarray}
\psi(x;z)&=&A(\rho,\pi) u_{\pi}^{-x} w_{\rho}^{-z},\;\;x > z. \nonumber\\ 
\psi(x;z)&=&A(\pi,\rho) u_{\pi}^{-x} w_{\rho}^{-z},\;\;x < z. \label{eq:apBA2pansatz} 
\end{eqnarray}
From this Ansatz, in which we have introduced the auxiliary labels $\pi$
and $\rho$ for later convenience, follow the eigenvalue and the relation between the two amplitudes before and after the collision
\begin{eqnarray}
\Lambda=u_{\pi},\;\;A(\rho,\pi)=S_{\l\s}(\rho,\pi)A(\pi,\rho).\nonumber\\
S_{\l\s}(\rho,\pi)=\left(\e^{\phi}u_{\pi}+\e^{\eta-\phi}u_{\pi}^{-1}w_{\rho}^{-1}\right).\label{eq:apBAslscat}
\end{eqnarray}
If we take periodic boundary conditions we get the following equations
for the complex numbers $u_{\pi}$ and $w_{\rho}$:
\begin{eqnarray}
u_{\pi}^N={A(\rho,\pi)\over
  A(\pi,\rho)}=S_{\l\s}(\rho,\pi).\nonumber\\
(u_{\pi}w_{\rho})^{N}=1.
\end{eqnarray}

One can check that for arbitrary $n_\l$ and $2n_\s=1$, the eigenvector is
given by
\begin{equation}
\psi(x;z_1,\dots,z_{n_\l})=\sum_{\rho}A(\Gamma)u_{\pi}^{-x}\prod_{k=1}^{n_\l}w_{\rho_k}^{-z_k},\;\;\Lambda=u_{\pi}.
\end{equation}
Here we have introduced a shorthand notation $\Gamma$ for the dependency of the
amplitude on the permutation $\rho$ and the order of types of domain
walls. This information is coded in the
following way. Let {\bf r} be the vector of coordinates $x,z_k$ of all
domain walls, ordered so that $r_m < r_{m+1}$.
The entries of the vector $\Gamma$ are the elements of the permutation
$\rho$ together with the auxiliary label $\pi$. The order of succession
in $\Gamma$ of $\pi$ and elements taken from $\rho$ matches that of $x$
and the $z_k$'s in {\bf r}, like in equation (\ref{eq:apBA2pansatz})

Amplitudes differing in an exchange of neighboring
s- and l-walls or an exchange of two neighboring l-walls are
related by the scattering matrix elements $S_{\l\s}(\rho_k,\pi)$
(\ref{eq:apBAslscat}) and $S_{\l\l}(\rho_k,\rho_l)$ respectively, where
$S_{\l\l}(\rho_k,\rho_l)=-1$ for $k\neq l$.
Imposing periodic boundary conditions gives us the BAE for these sectors:
\begin{eqnarray}
u_{\pi}^N&=&{A(\rho_1,\dots,\rho_{n_\l},\pi)\over
  A(\pi,\rho_1,\dots,\rho_{n_\l})}=\prod_{k=1}^{n_\l}S_{\l\s}(k,\pi).\nonumber\\ 
w_k^{-N}&=&S_{\l\s}(k,\pi)\prod_{j\neq k=1}^{n_\l}S_{\l\l}(k,j)=(-)^{n_\l-1}S_{\l\s}(k,\pi).
\end{eqnarray}

\subsection{Sectors with arbitrary values of $n_\s$ and $n_\l$}
Let $z$ be the coordinate of the l-wall and $x_i$ the
coordinates of the $i$th s-walls. By letting $t$ be the vertical
coordinate and $c_i$ the number of l-walls to the left of the $i$th
s-wall, we see that $(x_i+t+c_i\bmod 2)$ is a conserved quantity for
every s-wall. This means that the s-walls
split up into two kinds: odd and even ones. We shall therefore change
our notation a little bit: $x_i$ and $y_j$ will denote the coordinate
of the $i$th odd and the $j$th even s-wall respectively. Their total
numbers will be denoted by $n_\o$ and $n_\e$ and
$2n_\s=n_\o+n_\e$ and we shall use the shorthand $S_{\o,\e}$
for $S_{\s_{\o},\s_{\e}}$.

The Ansatz for the eigenvector in a general
sector is given by: 
\begin{eqnarray}
\psi(x_1,\dots,x_{n_\o};y_1,\dots,y_{n_\e};z_1,\dots,z_{n_\l})=\nonumber\\
\sum_{\pi}\sum_{\mu}\sum_{\rho}A(\Gamma)\prod_{i=1}^{n_\o}u_{\pi_i}^{-x_i}\prod_{j=1}^{n_\e}v_{\mu_j}^{-y_j}\prod_{k=1}^{n_\l}w_{\rho_k}^{-z_k},\label{eq:apBAvector}
\end{eqnarray}
with eigenvalue
\begin{equation}
\Lambda=\prod_{i=1}^{n_\o}u_i\prod_{j=1}^{n_\e}v_j.
\end{equation}
It turns out that in higher sectors all relations factorize into 
two-domain wall scattering relations, like (\ref{eq:apBAslscat}).
The BAE for $u_i$, $v_j$ and $w_k$ using periodic boundary conditions
then are
\begin{equation}
\begin{array}{rcl}
\dps u_i^{N} &=& \dps\prod_{m\neq
  i=1}^{n_\o}S_{\o,\o}(i,m)\prod_{j=1}^{n_\e}S_{\o,\e}(i,j)\prod_{k=1}^{n_\l}S_{\l\o}(k,i).\\ 
\dps v_j^{N} &=& \dps\prod_{i=1}^{n_\o}S_{\o,\e}(i,j)\prod_{m\neq
  j=1}^{n_\e}S_{\e,\e}(j,m)\prod_{k=1}^{n_\l}S_{\l\e}(k,j).\\
\dps w_k^{-N} &=& \dps\prod_{i=1}^{n_\o}S_{\l\o}(k,i)\prod_{j=1}^{n_\e}S_{\l\e}(k,j)\prod_{m\neq
  k=1}^{n_\l}S_{\l\l}(k,m).\end{array}\label{eq:apBABAE}
\end{equation}
Here, $S_{\l\o}$ is given by (\ref{eq:apBAslscat}) as well as
$S_{\l\e}$ with $u$ replaced by $v$. Furthermore, 
\begin{equation}
S_{\l\l}=S_{\o,\o}=S_{\e,\e}=-1,\qquad S_{\o,\e}=1.
\end{equation}

\section{Calculation of the function $G$}
\label{ap:calcG}
In the following we will always assume that the two curves have the
same limit point. In this case the Riemann surface of the function $G$
becomes disconnected and consists of an infinite number of discrete
parts comprising four sheets each. Each of these parts can be mapped
onto the plane by (\ref{eq:Tltdef}). We fix a particular part and
sheet of the Riemann surface by choosing $G$ equal to $f_\l$ at $z=0$,
on the sheet with the point $t=\e^{\i\pi/4}$. The form of $G$ on the
other sheets of this part of the Riemann surface is then
determined by the monodromy operators $\Gamma_\s$ and $\Gamma_\l$. The
finiteness of the densities $Q_\s$ and $Q_\l$ implies that the forms
$f_\s(z)\d z$ and $f_\l(z)\d z$ are non-divergent at $z=b$ and $z=b^*$. The
only singularities therefore are the simple poles at $z=0$ and
$z=\infty$. The residues of $G\d z$ at these points can be easily
calculated from $f_\l$ and its monodromy using equation
(\ref{eq:Tlmonop}). They are listed in Table \ref{ta:apCGtable1}.
From this table we see that the functions $f_\l$, $(1+Q_\l)f_\s-f_\l$, $-f_\l$ and
$-(1+Q_\l)f_\s+f_\l$ belong to the different sheets of the same part of the
Riemann surface of the function $G$. 
\begin{table}[hc]
\begin{center}
\begin{tabular}[c]{||l|l|l|l||}\hline
$z$ & $t_n$ & $G$ & $r_n$ \\ \hline
0 & $\e^{\i\pi/4}$ & $f_\l$ & 1\\
$\infty$ & $\i\e^{-\i\gamma/2}$ & $(1+Q_\l)f_\s+f_\l$ &
$Q_\l+2Q_\s-2$\\
0 & $-\e^{-\i\pi/4}$ & $(1+Q_\l)f_\s-f_\l$ & $Q_\l$\\
$\infty$ & $-\e^{-\i\gamma/2}$ & $-f_\l$ & $1-2Q_\s$\\
0 & $-\e^{\i\pi/4}$ & $-f_\l$ & $-1$\\
$\infty$ & $-\i\e^{-\i\gamma/2}$ & $-(1+Q_\l)f_\s-f_\l$ & $2-Q_\l-2Q_\s$\\
0 & $\e^{-\i\pi/4}$ & $-(1+Q_\l)f_\s+f_\l$ & $-Q_\l$\\
$\infty$ & $\e^{-\i\gamma/2}$ & $f_\l$ & $2Q_\s-1$\\\hline
\end{tabular}\end{center}
\caption{Poles and residues of $G\d z$}
\label{ta:apCGtable1}
\end{table}\\
The form $G\d z$ is defined
unambiguously by its residues and equals 
\begin{eqnarray}
G\d z&=&\sum_{n=1}^8 {r_n\over t-t_n}\d t\nonumber\\
&=&\left({1\over
2}\left(\e^{-\i\pi/4}t+\e^{\i\pi/4}t^{-1}\right)-{Q_l\over
2}\left(\e^{\i\pi/4}t+\e^{-\i\pi/4}t^{-1}\right)\right.+\nonumber\\
&&\hphantom{\left(\vphantom{1\over
2}\right.}\left.+C\left(t+t^{-3}\right)+C^*\left(t^{-1}+t^3\right)\vphantom{1\over
2}\right){1\over z(t)}{\d z(t)\over \d t}\d t.\label{eq:apCGG}
\end{eqnarray}
where $C$ given by
\begin{eqnarray} 
C&=&{1\over
  2\left(1+\e^{2\i\gamma}\right)}\left(-(1-Q_l)\left(\i\e^{\i\gamma/2}+\e^{\i\pi/4}\right){\vphantom{\pi\over
    4}}\right.\nonumber\\  
&&\hphantom{{1\over 2\left(1+\e^{2\i\gamma}\right)}}+\left.(1-2Q_s)(1-\i)\e^{\i\gamma/2}+2\i\sin{\pi\over 4}\right).
\end{eqnarray}
Or equivalently, one can write the domain wall densities as functions
of $\gamma$ and $C$:
\begin{eqnarray}
2Q_\s&=&1-\cos\left({\pi\over 4}+{\gamma\over
  2}\right)+Q_\l\cos\left({\pi\over 4}-{\gamma\over
  2}\right)\nonumber\\
&&{}-2\Re(C)\left(\cos{\gamma\over 2}+\cos{3\gamma\over
  2}\right)+2\Im(C)\left(\sin{3\gamma\over 2}-\sin{\gamma\over 2}\right).\label{eq:apCGQs}\\
Q_\l&=&\left(1+\sqrt{2}\cos{\gamma\over
  2}\right)^{-1}\left(1-\sqrt{2}\sin{\gamma\over
  2}\right.\nonumber\\
&&\hphantom{\left(1\right.}
\left.{}+2\Re(C)\left(\cos{3\gamma\over 2}+\cos{\gamma\over
  2}+\sin{3\gamma\over 2}-\sin{\gamma\over 2}\right)\right.\nonumber\\
&&\hphantom{\left(1\right.}
\left.{}+2\Im(C)\left(\cos{3\gamma\over 2}+\cos{\gamma\over
  2}-\sin{3\gamma\over 2}+\sin{\gamma\over 2}\right)\right).\label{eq:apCGQl}
\end{eqnarray}
The monodromy properties of the curves $\Xi$ and
$\Psi$ only hold when $b_1=b_2$. These curves are defined by the
form $G\d z$ taking only purely imaginary values on the tangential
vectors of $\Xi$ and $\Psi$, see equations (\ref{eq:TlXidef}) and
(\ref{eq:TlPsidef}). Given the initial conditions these
solutions must be unique except precisely at the points
$t=0$ and $t=\infty$. At these points the different solutions should
meet. If this were not the case the curves would intersect or would
not close, destroying the assumed monodromy assumption. The condition
$\Re(G\d z)=0$ must therefore be satisfied identically at the points
$t=0$ and $t=\infty$. From (\ref{eq:apCGG}) and (\ref{eq:Tltdef}) it
follows that 
\begin{equation}\begin{array}{l}
\dps \left. G\frac{\d z}{\d t}\d t\right|_{t=0} = \left. 4C(1+\e^{2\i\gamma})\d t\right|_{t=0}\\[4mm]
\dps \left. G\frac{\d z}{\d t}\d t\right|_{t=\infty} =
  \left. -4C^*(1+\e^{-2\i\gamma})\frac{\d t}{t^2}\right|_{t=\infty}  
\end{array}\end{equation}
The monodromy assumption translates itself therefore in the
condition $C=0$. The Riemann surface of $G$ then has infinitely
many disconnected parts with four sheets each. If $b_1\neq b_2$ these
parts become joined by regions of size $|b_1-b_2|$. One can still use
the variable $t$ with $b=(b_1+b_2)/2$. The function $G$ then is
single-valued except for the regions around $t=0$ and $t=\infty$ where
$|t|\sim |b_1-b_2|^{1/4}$ or $|t|^{-1}\sim |b_1-b_2|^{1/4}$
respectively. Equation (\ref{eq:apCGG}) gives the corrections to the
changes in the values of the residues if $C\neq 0$. It does not
however take into account contributions arising from the changes in
the topology of the Riemann surface. For small $t$ the form $G\d z$ is
approximated by terms of the order of $t^2\d t$ and $C\d t$. As a
lowest order approximation (\ref{eq:apCGG}) thus is valid if
$|t|^2\sim |b_1-b_2|^{1/2} < C$.

\section{Calculation of $\Lambda$}
\label{ap:calclamb}
In order to remove singularities at $z=0$ and $z=\infty$ occurring in
some of the expressions in this appendix, we introduce the following forms: 
\begin{equation}\begin{array}{rcl}
g_\l\d z&=&\dps \left({r_4 \over t-t_4}+{r_8\over t-t_8}\right)\d t\\[4mm]
g_\s\d z&=&\dps \left({r_2 \over t-t_2}+{r_6\over t-t_6}\right)\d t\\[4mm]
h_\l\d z&=&\dps \left({r_1 \over t-t_1}+{r_5\over t-t_5}\right)\d t\\[4mm]
h_\s\d z&=&\dps \left({r_3 \over t-t_3}+{r_7\over t-t_7}\right)\d t
\end{array}\end{equation}
The BAE (\ref{eq:BAlogba}) allow us to evaluate the
following expression using the definition $F_\l$ as
given by equation (\ref{eq:BAFldef}).
\begin{eqnarray}
 J_{\l,\infty} &=& \dps \Re\int_b^{\infty}\left(f_\l-g_\l\right)\d
 z\nonumber\\
&=& \dps \Re \lim_{t \to t_8}\left( F_\l(z(t))-r_4\log (t-t_4)-r_8\log
(t-t_8) \right)\nonumber\\
&=& \dps (1-2Q_\s)\log{|b||\cos\gamma|\over 4}+N^{-1}\log\Lambda +
2(1-Q_\s)\phi-\eta.\label{eq:apclintlinfzplane} 
\end{eqnarray}
The equality sign on the second line follows from the fact that
$F_\l(z)$ is a primitive function of $f_\l(z)$ and that $b$ is a
solution of the BAE (\ref{eq:BAlogba}). The expression on the third
line follows from the definition of $F_\l(z)$ as given by equation
(\ref{eq:BAFldef}). Similarly we find using also the definition for
$F_\s(z)$, equation (\ref{eq:BAFsdef}),
\begin{eqnarray}
J_{\s,\infty} &=& \Re\int_b^{\infty}\left((1+Q_\l)f_\s +
f_\l-g_\s\right)\d z\nonumber\\
&=& (2-Q_\l-2Q_\s)\log{|b||\cos\gamma|\over 4} +
N^{-1}\log\Lambda\nonumber\\
&&\hphantom{(2-Q_\l-2Q_\s)\log{|b||\cos\gamma|\over 4}} +
2(1-Q_\l-Q_\s)\phi-\eta.\label{eq:apclintsinfzplane}\\
J_{\l,0} &=& \Re\int_b^0\left(f_\l-h_\l\right)\d z\nonumber\\
&=& \log{4|b|\over |\cos\gamma|} - N^{-1}\log\Lambda +
2(1-Q_\s)\phi-\eta.\label{eq:apclintlzerozplane}\\
J_{\s,0} &=& \Re\int_b^0\left((1+Q_\l)f_\s-f_\l-h_\s\right)\d z\nonumber\\
&=& \dps Q_\l\log{4|b|\over |\cos\gamma|} +
N^{-1}\log\Lambda-2s_\l-2(1+Q_\l-Q_\s)\phi+\eta.\label{eq:apclintszerozplane}
\end{eqnarray}
with
\begin{equation}
s_\l=N^{-1}\sum_{j=1}^{n_\l}\log\psi_j.
\end{equation}
Eliminating $|b|$, $\phi$ and $s_\l$ from the
equations~(\ref{eq:apclintlinfzplane})$-$(\ref{eq:apclintszerozplane})
we find for the eigenvalue $\Lambda$ and the chemical potential $\eta$
\begin{eqnarray}
N^{-1}\log\Lambda-\eta {Q_\s Q_\l\over
  1-Q_\s}&=&Q_\s J_{\s,\infty}+{1\over 2}\left(1-2Q_\s+{Q_\s Q_\l\over
  1-Q_\s}\right)J_{\l,\infty}\nonumber\\[4mm]
&&\hspace{-2cm}-{1\over 2}\left(1-{Q_\s Q_\l\over
  1-Q_\s}\right)J_{\l,0}-\left(1-{Q_\s Q_\l\over 1-Q_\s}\right)\log{\cos\gamma\over 4}. 
\label{eq:apcleigval}
\end{eqnarray}
On the other hand, the integrals $J_{\l,\infty}$ and $J_{\s,\infty}$ can
easily be integrated in the t-plane using the results of appendix
\ref{ap:calcG}.
\begin{eqnarray}
J_{\l,\infty}&=&\dps \sum_{n\neq 4,8}^{8} r_n\log\left|t_n-t_8\right|.\\ 
J_{\s,\infty}&=&\dps \sum_{n\neq 2,6}^{8} r_n\log\left|t_n-t_2\right|.
\end{eqnarray}
With the help of Table \ref{ta:apCGtable1} of appendix \ref{ap:calcG} these
expressions are given explicitly by
\begin{eqnarray}
J_{\l,\infty}&=& {1\over 2}\log\left| {1-\cos\left({\pi/
    4}+{\gamma/ 2}\right)\over 1+\cos\left({\pi/
    4}+{\gamma/ 2}\right)}\right|+{Q_\l\over
  2}\log\left|{1+\cos\left({\pi/ 4}-{\gamma/ 2}\right)\over
  1-\cos\left({\pi/ 4}-{\gamma/ 2}\right)}\right|.
\label{eq:apclintlinftplane}\\[4mm]
J_{\s,\infty} &=& {1\over 2}\log\left| {1-\cos\left({\pi/
    4}-{\gamma/ 2}\right)\over 1+\cos\left({\pi/
    4}-{\gamma/ 2}\right)}\right|+{Q_\l\over
  2}\log\left|{1-\cos\left({\pi/ 4}+{\gamma/ 2}\right)\over
  1+\cos\left({\pi/ 4}+{\gamma/ 2}\right)}\right|.
\label{eq:apclintsinftplane}
\end{eqnarray}
And similarly $J_{\l,0}$ and $J_{\s,0}$ give:
\begin{eqnarray}
J_{\l,0} &=& {1\over 2}(2-Q_\l-2Q_\s)\log\left| {1+\cos\left({\pi/
    4}-{\gamma/ 2}\right)\over 1-\cos\left({\pi/
    4}-{\gamma/ 2}\right)}\right|\nonumber\\[4mm]
&&\hphantom{{1\over 2}(2-Q_\l-2Q_\s)}+{1\over
  2}(1-2Q_\s)\log\left|{1+\cos\left({\pi/ 4}+{\gamma/
    2}\right)\over 1-\cos\left({\pi/ 4}+{\gamma/
    2}\right)}\right|.
\label{eq:apclintlnultplane}\\[4mm] 
J_{\s,0} &=& {1\over 2}(2-Q_\l-2Q_\s)\log\left| {1+\cos\left({\pi/
    4}+{\gamma/ 2}\right)\over 1-\cos\left({\pi/
    4}+{\gamma/ 2}\right)}\right|\\[4mm] 
&&\hphantom{{1\over 2}(2-Q_\l-2Q_\s)}+{1\over
  2}(1-2Q_\s)\log\left|{1-\cos\left({\pi/ 4}-{\gamma/
    2}\right)\over 1+\cos\left({\pi/ 4}-{\gamma/ 2}\right)}\right|. 
\label{eq:apclintsnultplane}
\end{eqnarray}
Combining (\ref{eq:apclintlinftplane})$-$(\ref{eq:apclintsnultplane}) and
(\ref{eq:apcleigval}) then gives the eigenvalue expressed in the
densities (given by (\ref{eq:apCGQs}) and (\ref{eq:apCGQl})), the chemical
potential $\eta$ and the parameter $\gamma$.

\end{document}